\documentclass[conference]{IEEEtran}
\IEEEoverridecommandlockouts
\usepackage{amsmath,amssymb,amsfonts}
\usepackage{algorithmic}
\usepackage{graphicx}
\usepackage{textcomp}
\usepackage{xcolor}
\usepackage{lipsum}
\usepackage[sorting = none, backend = bibtex, style=numeric-comp]{biblatex}
\def\BibTeX{{\rm B\kern-.05em{\sc i\kern-.025em b}\kern-.08em
    T\kern-.1667em\lower.7ex\hbox{E}\kern-.125emX}}
\begin{document}


\textbf{IEEE Copyright Notice}\\\\

\begin{minipage}{\textwidth}
    {\copyright} {\copyright} \hspace{0.01mm} 2024 IEEE. Personal use of this material is permitted. Permission from IEEE must be obtained for all other uses, in any current or future media, including reprinting/republishing this material for advertising or promotional purposes, creating new collective works, for resale or redistribution to servers or lists, or reuse of any copyrighted component of this work in other works.\\

    
 \end{minipage}

\clearpage
\thispagestyle{empty}

\title{C2P-GCN: Cell-to-Patch Graph Convolutional Network for Colorectal Cancer Grading\\
}
\author{
    \IEEEauthorblockN{Sudipta Paul}
    \IEEEauthorblockA{\textit{Department of Electrical Engineering} \\
    \textit{Rensselaer Polytechnic Institute}\\
    Troy, New York, USA \\
    pauls5@rpi.edu}
    \and
    \IEEEauthorblockN{B\"{u}lent Yener}
    \IEEEauthorblockA{\textit{Department of Computer Science} \\
    \textit{Rensselaer Polytechnic Institute}\\
    Troy, New York, USA \\
    yener@cs.rpi.edu}
    \and
    \IEEEauthorblockN{Amanda W. Lund}
    \IEEEauthorblockA{\textit{Department of Pathology} \\
    \textit{NYU Grossman School of Medicine}\\
    New York, NY, USA \\
    amanda.lund@nyulangone.org}
}
\maketitle
\begin{abstract}
Graph-based learning approaches, due to their ability to encode tissue/organ structure information, are increasingly favored for grading colorectal cancer histology images. Recent graph-based techniques involve dividing whole slide images (WSIs) into smaller or medium-sized patches, and then building graphs on each patch for direct use in training.
This method, however, fails to capture the tissue structure information present in an entire WSI and relies on training from a significantly large dataset of image patches. In this paper, we propose a novel cell-to-patch graph convolutional network (C2P-GCN), which is a two-stage graph formation-based approach. In the first stage, it forms a patch-level graph based on the cell organization on each patch of a WSI. In the second stage, it forms an image-level graph based on a similarity measure between patches of a WSI considering each patch as a node of a graph. This graph representation is then fed into a multi-layer GCN-based classification network. Our approach, through its dual-phase graph construction, effectively gathers local structural details from individual patches and establishes a meaningful connection among all patches across a WSI. As C2P-GCN integrates the structural data of an entire WSI into a single graph, it allows our model to work with significantly fewer training data compared to the latest models for colorectal cancer.  Experimental validation of C2P-GCN on two distinct colorectal cancer datasets demonstrates the effectiveness of our method. 
\end{abstract}
\section{Introduction}
Recent advancements in deep learning techniques have demonstrated higher efficacy in diagnosing colorectal cancer, outperforming traditional hand-crafted machine learning techniques. Numerous convolutional neural network (CNN)-based architectures \cite{extend, china} were introduced in recent years to automatically grade colorectal histology images. These methods mostly break larger-size images into smaller patches, which are then used for both training and making predictions. 
Unfortunately, these methods fail to capture the tissue structure information as the features obtained from smaller patch sizes might not have a clear interpretative connection with the glandular architecture presented in colorectal cancer images. 

To alleviate the issues mentioned above, recently some graph convolutional network (GCN) based architectures \cite{cgcnet, hatnet, HACT, vasu} were introduced which utilize the tissue structure information through the cell graphs technique \cite{cg}. 
For the colorectal cancer grading task, GCN-based architectures such as CGC-Net \cite{cgcnet} and HAT-Net \cite{hatnet} demonstrated notably better performance compared to other methods. However, similar to CNN-based methods, both CGC-Net and HAT-Net divide the large CRC histology images into medium-sized patches and construct cell graphs on those. The graphs constructed on each patch are then used for training, requiring a substantially large amount of training data. 
While these methods capture the tissue structure information on each patch, they do not establish a discernible interpretive connection across the whole image.

In this paper, we propose a novel Cell-to-Patch Graph Convolution Network (C2P-GCN) that features a two-stage graph construction process; a patch-level graph, and an image-level graph. The process is initiated by breaking a WSI into smaller or medium-sized patches. Then, at the patch level, a cell graph alongside various well-known global graphs (Voronoi graph of cells) \cite{gg} is created for each patch. Next, an image-level graph is constructed which is built over the entirety of a large image or a WSI, treating each patch as a node. The edges (connections) between these nodes (patches) are formed by a similarity measure between patches, thereby mapping out an interpretive connection among similar patches throughout the whole image. By encoding the entire image data into a unified graph, C2P-GCN processes it through a sophisticatedly designed multi-layer GCN-based classification network.

The key contributions of this paper are: (a) We introduced a novel C2P-GCN architecture, a dual-stage graph construction-based approach that harnesses the tissue structure information from the local cellular organization via patch-level graphs; then forms a meaningful connection among patches within a WSI that exhibit similar characteristics using an image-level graph; (b) We performed comprehensive experiments on two distinct colorectal cancer dataset, Extended CRC dataset \cite{extend} and the Colon Cancer Dataset from Zhejiang University in China \cite{china}, showing that C2P-GCN exhibiting strong performance on par or even surpassing that of the recently proposed CNN and GCN-based methods. (c) Since C2P-GCN incorporates the structural details from an entire WSI into a single graph, it allows our model to require considerably less training data. On our tested datasets, C2P-GCN uses over two orders of magnitude less training data compared to the state-of-the-art methods while yielding comparable or better results.



\begin{figure*}[!t]
\centering
\includegraphics[width=18cm, height=8.5cm]{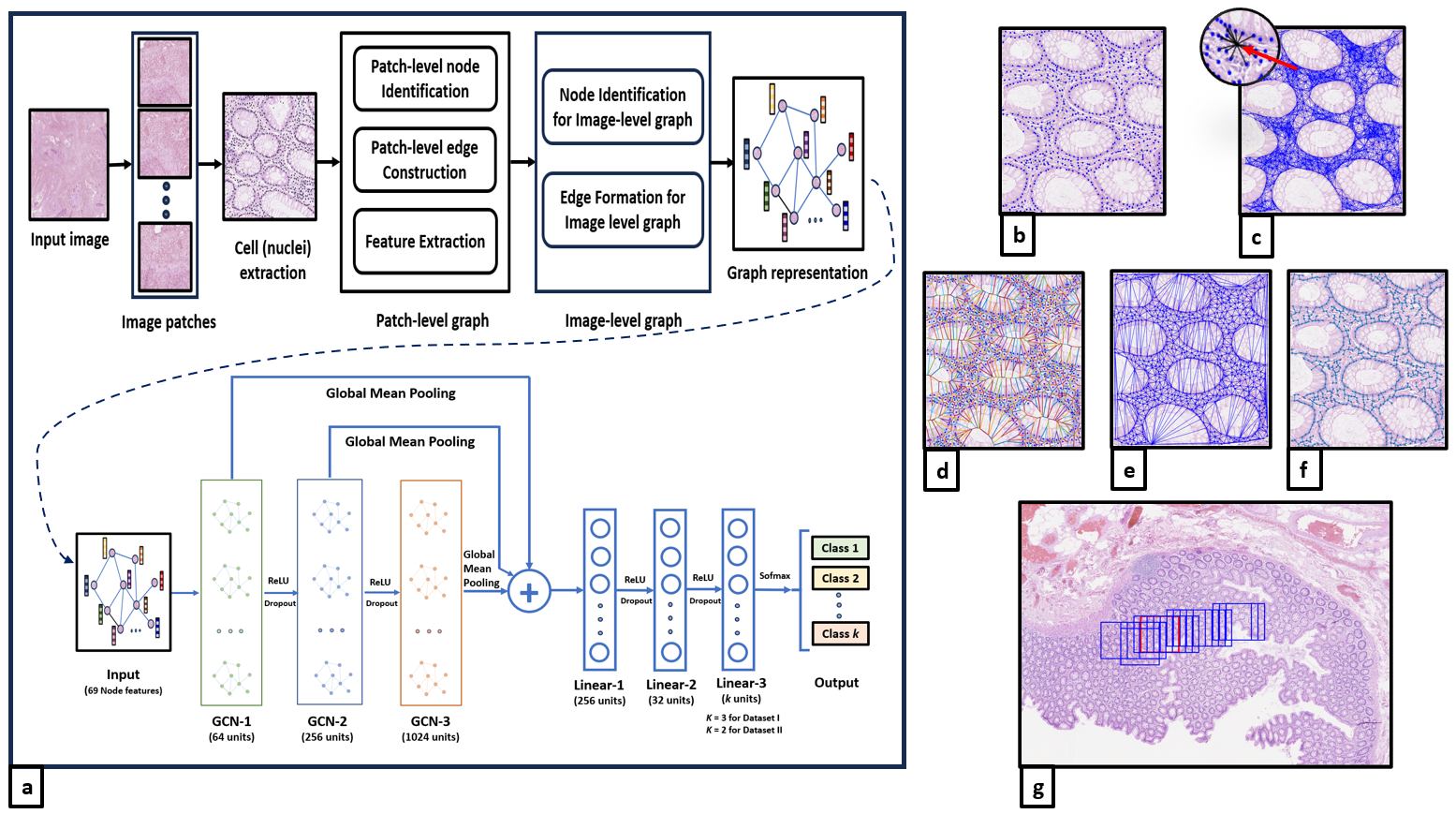}
\caption{(a) C2P-GCN overall pipeline. C2P-GCN initially breaks a WSI into multiple patches and then constructs a patch-level graph for capturing structural features within a patch. Next, it forms an image-level graph of collective patches based on similar measures. The image-level graph containing whole image structural information is fed into a multi-layer GCN structure for classification. (b) Nuclei detection with gLoG filter. (c) Cell-graph construction; the magnified section shows how a sample node (nuclei) is connected with its neighborhood. (d) Voronoi diagram. (e) Delaunay triangulation. (f) Minimum spanning tree. (g) This figure highlights the red patch, which is a randomly chosen node of interest, and its top 15 most similar patches (blue) in terms of similarity scores.}
\label{fig:your_label}
\end{figure*}

\section{Methodology}

A complete pipeline of the methodologies we propose in this paper is depicted in Fig. 1(a), which involves dividing a WSI into smaller or medium-sized patches, constructing patch-level and image-level graphs, and finally feeding these graphs into a multi-layer GCN for classification.



\subsection{Cell identification}

Once a WSI is divided into patches, the next task is to extract cells from each patch to construct patch-level graphs in the following step. We note that the method
proposed in this paper does not require precise cell segmentation. Determining
only the cell locations is enough to form a graph or extract necessary features. In this work, we
adopted the generalized Laplacian of Gaussian (gLoG) filter-based automatic nuclei detection technique developed in \cite{glog}. 
An illustrative example of nuclei detection with a gLoG filter on a colorectal cancer image patch is presented in Fig. 1(b). Table I shows the gLoG kernel parameters and the corresponding values we used to detect nuclei from the colorectal cancer images. 
\vspace{-2mm}
\begin{table}[!htbp]
\caption{kernel parameters}
\label{table_example}
\centering
\begin{tabular}{|c|c|c|c|}
\hline
x-axis scale, $\sigma_x$ & y-axis scale, $\sigma_y$ & Orientation, $k$ & Bandwidth, $w$ \\
\hline
8 & 4 & 9 & 7 \\
\hline
\end{tabular}
\end{table}
\subsection{Patch graph formation} 

\subsubsection{Cell graph} 

Let, $G_p = (V_p, E_p)$ be the cell graph constructed on a patch where $V_p$ and $E_p$ denote the set of nodes and edges of the graph. We consider the cells extracted from the previous stage as the node set $V_p$ for our cell graph. An edge $(u, v)$ between two nodes (cells) $u$ and $v$ is determined based on biological understanding and the known interactions between cells in a particular tissue type. We propose that cells closer together in Euclidean distance are more prone to interact. Therefore, an edge is assigned between two nuclei if they are within a predefined distance $d_p$ from each other. From this, the adjacency matrix is constructed as:

\vspace{-1mm}

\begin{table}[!b]
\renewcommand{\arraystretch}{1.3}
\caption{Cell graph features}
\label{table_example}
\centering
\begin{tabular}{@{}|p{2cm}|c|p{5cm}|@{}}
\hline
\textbf{Feature type} & \textbf{No.} & \textbf{Description} \\ \hline
Connectedness and cliquishness measures & 4 & Average degree; Clustering coefficient; Giant connected component ratio; Number of connected components \\ \hline
Distance-based measures & 8 & Average eccentricity, Diameter, Radius, Average path length, Number of central points, Percent of central points, Number of vertices, and Number of edges \\ \hline
Spectral Measures & 6 & Largest eigenvalue adjacency, Trace of adjacency, Energy of adjacency, Lower slope, Upper slope, Trace of Laplacian \\ \hline
\end{tabular}
\end{table}
$$
A_p(i, j)=\left\{\begin{array}{cc}
1 & \text { if } D_p(i, j)<d_p \\
0 & \text { otherwise }
\end{array}\right.
$$
Here, the value of $d_p$ is set to be $64$. An example of this graph built within a patch is illustrated in Fig. 1(c). The cell graph constructed on each patch made it possible to assess and quantify the interactions occurring within localized neighborhoods. A total of $18$ distinct cell graph features are extracted to characterize this local cell organization within each patch, as shown in Table II \cite{cg}.

\subsubsection{Voronoi graphs of cells for global graph} 
We applied three distinct methods: the Voronoi Diagram (VD), Delaunay Triangulation (DT), and the Minimum Spanning Tree (MST) to construct global graphs in each patch, and illustrations of each of these global graphs applied to an individual patch are shown in Figure 1(d), 1(e), and 1(f), respectively. To quantify the global graph tesselations of cells present within an entire patch, $24$ distinct features ($12$ Voronoi, $8$ Delaunay, and $4$ MST) are extracted and are presented in Table III \cite{gg}. Apart from that, 27 nuclear density attributes were also integrated to characterize the nuclei clustering within the patch.



\begin{table}[!h]
\renewcommand{\arraystretch}{1.3}
\caption{Global graph features}
\label{table_example}
\centering
\begin{tabular}{@{}|p{1.75cm}|c|p{5.25cm}|@{}}
\hline
\textbf{Feature type} & \textbf{No.} & \textbf{Description} \\ \hline
Voronoi diagram & 12 & Chord length, polygon area, perimeter: mean, SD, min/max ratio, disorder \\ \hline
Delaunay triangulation & 8 & Triangle side length, area: mean, SD, min/max ratio, disorder \\ \hline
Minimum spanning tree & 4 & Edge length: mean, SD, min./max. ratio, disorder \\ \hline
Nuclei nearest neighbor (NN) features & 27 & Number of nuclei; area of polygons; density of nuclei; mean, SD, and disorder of distance to $k$-NN ($k=3,5,7$); mean, SD, and disorder of NN in a $n$ pixel radius ($n = 10,20,30,40,50$) \\ \hline
\end{tabular}
\end{table}
\subsection{Image level Graph}
For the image level graph, $G_I = (V_I, E_I)$, we treat each patch from a WSI as an individual node from the node set, $V_I$, with a $d$ dimensional feature vector, $x_i \in \mathbf{R}^d$ for $i \in V_I$. Here, the node attributes are the same features set we obtained from each patch in our patch-level graph, i,e; $d = 69$. $E_I$ represents the edge set and $e_{i,j} \in E_I$ denotes an edge between two patches. To form an edge between a pair of nodes (patches), we rely on assessing the similarity in attributes between these nodes, facilitating connections between nodes that exhibit similar traits. For this, we use cosine similarity as a similarity measure and form an edge between a pair of nodes if the similarity value exceeds a predefined threshold, $\theta$, and the edge weights are set as the similarity values. Mathematically, 
$$
w_{ij} = cos(i,j) = \dfrac{v_i \cdot v_j}{||v_i|| \hspace{1mm} ||v_j||}
$$where, $w_{ij}$ represents the edge weight between node $i$ and $j$. Now, the adjacency matrix for this graph is computed as 
$$
A_I(i, j)=\left\{\begin{array}{cc}
w_{ij} & \text { if } w_{ij}>\theta \\
0 & \text { otherwise }
\end{array}\right.
$$
The edges between these patches exhibiting similar characteristics establish a meaningful link across the whole image. Here, $\theta$ is chosen as $0.8$ which is obtained by manual tuning. A visual representation of a specific patch and its connections to other patches based on similarity is shown in Fig. 1(g).  
\subsection{Network architecture}
 In the C2P-GCN architecture, we use a multi-layer Graph Convolutional Network (GCN) followed by a series of linear layers and a softmax classification layer. For each GCN layer, $GCN_{l}$, with ReLU activation and dropout, the equations might be represented as follows:

\begin{equation}
    H^{l+1} = \text{Dropout}\left( \text{ReLU} \left({\text{GCN}}_l\left( X^{l}, A'_I; W^{l} \right) \right) \right)
\end{equation}
here, $A'_I$ is the weighted adjacency matrix where ${(A'_{ij})}_I = {(A_{ij})}_I \cdot w_{ij}$. The learnable weight matrix at layer $l$ is presented as $W^{l}$, the input to a GCN layer $l$ is $X^{l}$, and $H^{l+1}$ is the node feature matrix at layer $l+1$. We apply a global mean pooling on the feature representation of each layer and then concatenate the pooled features from all GCN layers $L$ as:
\begin{equation}\nonumber
P_{\text{concat}} = \text{Concatenate}\left(P^{(1)}, P^{(2)}, \ldots, P^{(L)}\right), {\text{where}}
\end{equation} 
\begin{equation}\nonumber
P^{(l)} = \text{GlobalMeanPool}\left( \text{GCN}_l\left(X^{l}, A'_I; W^{l} \right) \right)
\end{equation}
$P_{\text{concat}}$ is then passed through a series of linear layers with ReLU activations and dropout as

\begin{equation}
z^{(t+1)} = \text{Dropout}\left( \text{ReLU}\left( w^{(t)}_{\text{linear}} z^{(t)} + b^{(t)}_{\text{linear}} \right) \right)
\end{equation}
where, $z^{(1)} = P_{\text{concat}}$. In (2), the weights and biases for the linear layers are denoted as $w^{(t)}_{\text{linear}}$ and  $b^{(t)}_{\text{linear}}$ respectively, where the number of linear layers $t$ ranges from $1$ to  $N$. Finally, a softmax layer is used for classification as follows

\begin{equation}
 Y = \text{Softmax}\left( w^{(N+1)}_{\text{linear}} z^{(N)} + b^{(N+1)}_{\text{linear}} \right)   
\end{equation}
Multi-layer GCN architecture used in C2P-GCN along with the dimension of each layer is visualized in Fig. 1(a).

\section{Experiments}
\subsection{Dataset}
Our proposed method is evaluated on two different colorectal cancer datasets. Dataset I \cite{extend}, the Extended CRC dataset consists of $300$ images at $20$x magnification and contains $120$ normal, $120$ low grade, and $60$ high grade images, with the size of $4548$ $\times$ $7548$ and $5000$ $\times$ $7300$ pixels. We divide each large image into $768$ $\times$ $768$ image patches with a stride of $128$. To ensure fair comparisons with other methods \cite{hatnet, cgcnet, extend}, we divided the dataset into three equal folds for cross-validation, maintaining the identical split of the images in each fold. We then train our model with two folds at a time and validate it on another fold. 

Dataset II \cite{china} is from the Department of Pathology of Zhejiang University in China which consists of $717$ large images. It has $355$ cancer and $362$ normal images at $40$x magnification of various sizes. For a fair comparison with \cite{china}, we randomly split the images in such a way that half of them form the training set and the other half form the testing set while preserving the same class proportion. For this dataset, we divide each image into small patches of size $768$ $\times$ $768$ with a stride of $256$ for cell-graph construction.



\subsection{Implementation}

The C2P-GCN is implemented with the PyTorch framework, utilizing the PyTorch Geometric (PyG) library. 
In this experiment, we used the Adam optimizer with a learning rate of $0.0002$ selected through a grid search process. The batch size is chosen as $20$, the model is trained for $600$ epochs and dropout regularization is chosen as $0.3$.

\subsection{Experimental Results}

To validate our approach, we compared the performance of C2P-GCN with several state-of-the-art methods. As illustrated in Table IV, our method outperforms all the CNN-based methods, MobileNet, ResNet50, Inception, Xception, CA-CNN, and VIT by a large margin. It also outperforms the GCN-based method, CGC-Net comfortably. There is a small gap between our method and HAT-Net, which we believe could be mitigated by adding more training data. It is noted that CGC-Net and HAT-Net were both trained on image patches, then, a majority voting was applied to aggregate patch-level predictions into image-level classifications. For the Extended CRC dataset, \cite{hatnet} extracted a total of $114243$ patches from the large images, and used approximately two-thirds of the patches for training. With C2P-GCN, on the other hand, we trained our model on the image-level graph data, and only $200$ training data were used at a time to predict performance on each fold. This means C2P-GCN uses more than two orders of magnitude smaller training data than those used by the HAT-Net while experiencing a performance decrease of just $0.33\%$.

On Dataset II, we worked only on the binary classification since the dataset is highly imbalanced to conduct a multiclass classification with C2P-GCN as our approach leverages full images for training rather than individual patches. For binary classification, as shown in Table V, our method yields the highest performance compared to other methods implemented on this dataset and beats the SVM-CNN method by $0.4\%$. It is noted that our model was trained only with $359$ image-level graph data. The SVM-CNN methods did not directly mention the training size, however, Dataset II has an average image size of $5.10$ $mm^2$ $(10000 \times 10000)$, cropped at 40x magnification i.e. $226$ nm/pixel \cite{china}. Since the authors used a patch size of $672 \times 672$ to break each large image and used patches extracted from half of the total image dataset for training purposes, it is safe to say that, C2P-GCN uses over two orders of magnitude less training data compared to what is used by SVM-CNN.

\section{Conclusion}

In this paper, we introduced a novel cell-to-patch graph convolutional network (C2P-GCN) method, which effectively integrates the structural information of a whole slide image (WSI) into one comprehensive graph using a dual-phase graph formation strategy. 
Our method when applied to two different colorectal cancer datasets, not only yields comparable or better performance than most of the recently proposed CNN and GCN-based architectures but also does so with a substantially less amount of training data. C2P-GCN, exhibiting strong performance confirms the significance of our contribution.

\begin{table}[!t]
\caption{Results on Dataset I}
\label{table_example}
\centering
\begin{tabular}{|c|c|c|c|}
\hline
Methods & Accuracy (\%) & Methods & Accuracy (\%) \\ \hline
MobileNet \cite{extend} & 84.33 $\pm$ 3.30 & ResNet50 \cite{extend} & 84.33 $\pm$ 0.94 \\ \hline
Inception \cite{cgcnet} & 84.67 $\pm$ 1.70 & Xception \cite{hatnet} & 86.67 $\pm$ 0.94 \\ \hline
CA-CNN \cite{extend} & 86.67 $\pm$ 1.70 & VIT \cite{hatnet} & 86.67 $\pm$ 4.04 \\ \hline
CGC-Net \cite{cgcnet} & 93.33 $\pm$ 0.93 & HAT-Net \cite{hatnet} & 95.33 $\pm$ 0.58 \\ \hline
\multicolumn{2}{|c|}{{\bf Ours (C2P-GCN)}} & \multicolumn{2}{c|}{95.00 $\pm$ 1.70} \\ \hline
\end{tabular}
\end{table}
\begin{table}[t]
\caption{Results on Dataset II}
\label{table_example}
\centering
\begin{tabular}{|c|c|c|c|}
\hline
Methods & Accuracy & Methods & Accuracy \\ \hline
MCIL \cite{mcil}  & $95.5\%$ & TRANS \cite{trans} & $92.3\%$ \\ \hline
SVM-IMG \cite{china} & $94.3\%$ & SVM-MF \cite{china} & $90.1\%$ \\ \hline
SVM-CNN \cite{china} & $98.0\%$ & {\bf Ours (C2P-GCN)} & $98.4\%$ \\ \hline
\end{tabular}
\end{table}

\printbibliography

@article{china,
  title={Large scale tissue histopathology image classification, segmentation, and visualization via deep convolutional activation features},
  author={Y, Xu and Z, Jia and LB, Wang and others},
  journal={BMC Bioinformatics},
  volume={18(1):281},
  year={2017},
}

@ARTICLE{glog,
  author={Xu, Hongming and Lu, Cheng and Berendt, Richard and Jha, Naresh and Mandal, Mrinal},
  journal={IEEE Journal of Biomedical and Health Informatics}, 
  title={Automatic Nuclei Detection Based on Generalized Laplacian of Gaussian Filters}, 
  year={2017},
  volume={21},
  number={3},
  pages={826-837}}

@ARTICLE{extend,
  author={Shaban, Muhammad and Awan, Ruqayya and Fraz, Muhammad Moazam and Azam, Ayesha and Tsang, Yee-Wah and Snead, David and Rajpoot, Nasir M.},
  journal={IEEE Transactions on Medical Imaging}, 
  title={Context-Aware Convolutional Neural Network for Grading of Colorectal Cancer Histology Images}, 
  year={2020},
  volume={39},
  number={7},
  pages={2395-2405},}

@inproceedings{hatnet,
  title={HAT-Net: A Hierarchical Transformer Graph Neural Network for Grading of Colorectal Cancer Histology Images},
  author={Yihan Su and Yutian Bai and Bo Zhang and Zheng Zhang and Wendong Wang},
  booktitle={BMVC},
  year={2021},}

@article{cgcnet,
  title={CGC-Net: Cell Graph Convolutional Network for Grading of Colorectal Cancer Histology Images},
  author={Yanning Zhou and Simon Graham and Navid Alemi Koohbanani and Muhammad Shaban and Pheng-Ann Heng and Nasir M. Rajpoot},
  journal={2019 IEEE/CVF International Conference on Computer Vision Workshop (ICCVW)},
  year={2019},
  pages={388-398},}

@article{mcil,
  title={Weakly supervised histopathology cancer image segmentation and classification.},
  author={Xu, Y. and Zhu, J.Y. and Chang, E.I. and Lai, M. and Tu, Z.},
  journal={MIA},
  year={2014},}

@article{gg,
  author={Lee, G. and Ali, S. and Epstein, JI. and  Madabhushi, A. and Veltri, RW},
  title={Nuclear Shape and Architecture in Benign Fields Predict Biochemical Recurrence in Prostate Cancer Patients Following Radical Prostatectomy: Preliminary Findings.},
  journal={Eur Urol Focus},
  year={2017},
  pages={457–466},
}

@article{trans,
  title={Discriminative data transform for image feature extraction and classification.},
  author={Song, Y. and Cai, W. and Huh, S. and Chen, M. and Kanade, T. and Zhou, Y. and Feng, D.},
  journal={MICCAI},
  year={2013},
}

@article{cg,
author = {Yener, B\"{u}lent},
title = {Cell-graphs: image-driven modeling of structure-function relationship},
year = {2016},
issue_date = {January 2017},
publisher = {Association for Computing Machinery},
address = {New York, NY, USA},
volume = {60},
number = {1},
issn = {0001-0782},
journal = {Commun. ACM},
month = {dec},
pages = {74–84},
numpages = {11},
}

@article{HACT,
  title={HACT-Net: A Hierarchical Cell-to-Tissue Graph Neural Network for Histopathological Image Classification},
  author={Pushpak Pati and Guillaume Jaume and Lauren Alisha Fernandes and Antonio Foncubierta and Florinda Feroce and Anna Maria Anniciello and G. Scognamiglio and Nadia Brancati and Daniele Riccio and Maurizio Do Bonito and Giuseppe De Pietro and Gerardo Botti and Orcun Goksel and Jean-Philippe Thiran and Maria Frucci and Maria Gabrani},
  journal={ArXiv},
  year={2020},
  volume={abs/2007.00584}
}

@ARTICLE{vasu,
  author={Acharya, Vasundhara and Choi, Diana and Yener, Bülent and Beamer, Gillian},
  journal={IEEE Access}, 
  title={Prediction of Tuberculosis From Lung Tissue Images of Diversity Outbred Mice Using Jump Knowledge Based Cell Graph Neural Network}, 
  year={2024},
  volume={12},
  number={},
  pages={17164-17194}}

\end{document}